\documentclass{ws-procs9x6}

\usepackage{epsfig}

\def\be{\begin{eqnarray}}
\def\ee{\end{eqnarray}}
\def\lsim{\mathrel{\rlap{\lower3pt\hbox{\hskip1pt$\sim$}}
     \raise1pt\hbox{$<$}}} 
\def\gsim{\mathrel{\rlap{\lower3pt\hbox{\hskip1pt$\sim$}}
     \raise1pt\hbox{$>$}}} 
\def\la{\langle}\def\ra{\rangle}

\def\cal{\it}

\begin{document}

\title{A new state of matter at high temperature as ``sticky molasses"}

\author{Gerald E. Brown
\footnote{\uppercase{I}nvited talk at the 
\uppercase{KIAS-APCTP S}ymposium in \uppercase{A}stro-\uppercase{H}adron
\uppercase{P}hysics ``\uppercase{C}ompact 
\uppercase{S}tars: \uppercase{Q}uest for \uppercase{N}ew 
\uppercase{S}tates of \uppercase{D}ense \uppercase{M}atter",
\uppercase{N}ovember 10-14, \uppercase{S}eoul, \uppercase{K}orea}
}

\address{Department of Physics and Astronomy,\\
          State University of New York, Stony Brook, NY 11794, USA }

\author{Chang-Hwan Lee}
\address{Department of Physics and\\
Nuclear Physics \& Radiation Technology Institute (NuRI),\\
Pusan National University,
              Pusan 609-735, Korea} 
\author{Mannque Rho}
\address{Service de Physique Th\'eorique, CEA/Saclay, 91191 Gif-sur-Yvette,
France,\\ Korea Institute for Advanced Study, Seoul
130-722, Korea \\
\& Department of Physics, Hanyang University, Seoul 133-791, Korea}

\maketitle

\abstracts{
The main objective of this work is to explore the evolution in the
structure of the quark-antiquark bound states in going down in the
chirally restored phase from the so-called ``zero binding points"
$T_{zb}$ to the QCD critical temperature $T_c$ at which the
Nambu-Goldstone and Wigner-Weyl modes meet. In doing this, we
adopt the idea recently introduced by Shuryak and Zahed for
charmed $\bar c c$, light-quark $\bar q q$ mesons
$\pi, \sigma, \rho, A_1$ and gluons that at $T_{zb}$,  the
quark-antiquark scattering length goes through $\infty$ at which
conformal invariance is restored, thereby transforming the matter
into a near perfect fluid behaving hydrodynamically, as found at
RHIC. We name this new state of matter as ``sticky molasses".
We show that the binding of these states is accomplished by
the combination of (i) the color Coulomb interaction,
(ii) the relativistic effects,
and (iii) the interaction induced by the instanton-anti-instanton
molecules. The spin-spin forces turned out to be small. While near
$T_{zb}$ all mesons are large-size nonrelativistic objects bound
by Coulomb attraction, near $T_c$ they get much more tightly
bound, with many-body collective interactions becoming important
and making the $\sigma$ and $\pi$ masses approach zero (in the chiral limit).
The wave function at the origin grows strongly with binding, and
the near-local four-Fermi interactions induced by the instanton
molecules play an increasingly more important role as the
temperature moves downward toward $T_c$.
}


\section{Introduction\label{intro}}

The concept that hadronic states may survive in the high
temperature phase of QCD, the quark-gluon plasma\cite{BLRS},
has been known for some time. In particular, it
 was explored by Brown et
al.\cite{BBP91,BJBP93}. The properties of (degenerate) $\pi$ and $\sigma$
resonances above $T_c$  in the context of the
NJL model was discussed earlier by Hatsuda
and Kunihiro\cite{Hatsuda_sigma},
 and in the instanton liquid model by Sch\"afer and Shuryak \cite{SS_survive}.
 Recently,
lattice calculations \cite{datta02,hatsuda2003} have shown
that, contrary to the original suggestion by Matsui and Satz  \cite{MS},
the lowest charmonium states $J/\psi,\eta_c$ remain bound well above $T_c$.
The estimates of the zero binding temperature for
charmonium
 $T_{J/\psi}$ is now limited to the interval
 $2T_c > T_{J/\psi} > 1.6 T_c$, where $T_c\approx 270\, MeV$
 is that for quenched QCD.
 Similar results for
 light quark mesons exist but are less quantitative at the moment.
However since the ``quasiparticle" masses close to $T_c$ are
large, they must be similar to  those for charmonium states.

In the chiral limit all states above the chiral restoration go into
chiral multiplets. For quark quasiparticles this is also true,
but although the chirality is conserved during their propagation,
they are not massless and move slowly near $T_c$ where their
``chiral mass'' $m=E(p\rightarrow 0)$ is large ($\sim 1$ GeV).

RHIC experiments have found that hot/dense matter at temperatures
above the critical value $T_c\approx 170 \, MeV$ is $not$ a weakly
interacting gas of quasiparticles, as was widely expected. We
envision it to be ``sticky molasses." Indeed, RHIC data have
demonstrated the existence of very robust collective flow
phenomena, well described by ideal hydrodynamics. Most decisive in
reaching this conclusion was the early measurement of the elliptic
flow which showed that equilibration in the new state of matter
above $T_c$ set in in a time $< 1$ fm/c \cite{hydro2001}.
Furthermore, the first viscosity estimates \cite{Teaney2003} show
surprisingly low values, suggesting that this matter is the most
perfect liquid known. Indeed,  the ratio of shear
  viscosity
coefficient to
  the entropy is only $\eta/s\sim 0.1$,  two orders of magnitude
  less than for water.
Furthermore, it is comparable to
predictions in the infinite coupling limit\cite{PSS}
(for $\cal N$=4 SUSY YM theory)  $\eta/s = 1/4\pi$, perhaps the
lowest value possible.

Shuryak and Zahed\cite{shuryak2003} (hereafter referred to as SZ
whenever unambiguous) have recently connected these two issues
together. They have
 suggested that large rescattering cross sections apparently present
 in hot matter at RHIC
 are generated by resonances near the zero-binding lines.
 Indeed,  at the point of
zero binding  the scattering
length $a$ of the two constituents goes to $\infty$ and this
provides
low viscosity. This phenomenon is analogous to the  elliptic flow
observed in the expansion of trapped $^6$Li atoms rendered
possible by tuning the scattering length to very large values via
a Feshbach resonance~\cite{Li6}.

Near the zero-binding points, to be denoted by $T_{zb}$, introduced by SZ
the binding is small and thus
the description of the system can be simple and nonrelativistic.
The binding comes about chiefly
from the attractive Coulomb color electric field, as evidenced in
lattice gauge calculation of Karsch and
collaborators\cite{datta02,petreczky02}, and Asakawa and
Hatsuda\cite{hatsuda2003}, as we shall detail. The instanton
molecule interactions, which we describe below, are less important
 at these high temperatures ($T \sim 400$ MeV).
All changes as one attempts (as we show below) to dicuss
the more deeply bound states just above $T_c$.

In another  work \cite{SZ_CFT}, Shuryak and Zahed have also found
sets of highly relativistic bound light states  in the strongly coupled
 $\cal N$=4 supersymmetric Yang-Mills theory at finite
temperature (already mentioned above in respect to viscosity).
They suggested that the very strong Coulomb attraction can be
balanced by high angular momentum, producing light states with
masses $m\sim T$. Furthermore, the density of such states remains
constant  at arbitrarily large  coupling. They argued that in this
theory a transition from weak to strong coupling basically implies
a smooth transition from a gas of quasiparticles to a gas of
``dimers'', without a phase transition. This is
an important part of the overall emerging
 picture, relating strong coupling, viscosity and light bound states.

In this work we wish to construct the link between the chirally
broken state of hadronic matter below $T_c$ and the chirally
restored mesonic, glueball state above $T_c$.
Our objective  is to understand and to
work out in detail what exactly happens with hadronic states
at temperatures between
$T_c$ and $T_{zb}$.
 One important new point
 is that these chirally restored
hadrons are so small that the color charges are
locked into the hadrons at such short distances ($< 0.5$ fm)
that the Debye screening
is unimportant. This is strictly true at $T \gsim T_c$,
 where there is very little free charge.
In this temperature range the nonrelativistic treatment of SZ should
be changed to a relativistic one.

The relativistic current-current interaction, ultimately related with
the classical Ampere law,
 is about as important as the Coulomb one,
effectively doubling the attraction
(see section \ref{sec_coul_rel}).  We also found that the spin-spin
forces discussed in \ref{sec_spinspin} are truly negligible.
In effect, with the help of the instanton molecule interaction,
one can get the bound quark-antiquark states down in energy, reaching
the massless $\sigma$ and $\pi$ at $T_c$,
 so that a smooth transition can be made with the chiral
breaking at $T<T_c$.

The non-pertubative interaction from the
instanton molecules becomes very important.
Let us remind the reader of the history of the issue.
  The nonperturbative gluon condensate, contributing to
the dilatational charge or trace of the stress tensor
$T_{\mu\mu}=\epsilon-3p$,
 is not melted at $T_c$. In fact more than half of the vacuum gluon condensate
 value remains at $T$ right above $T_c$.
 the hard glue or epoxy which
explicitly breaks scale invariance but is
 unconnected with hadronic masses. The rate at which the epoxy
is melted can be measured by lattice gauge simulations, and this
tells us the rate at which the instanton molecules are broken
up with increasing temperature\cite{BLRS}.

As argued by Ilgenfritz and Shuryak \cite{IS}, this phenomenon can
be explained by breaking of the instanton ensemble into instanton
molecules with zero topological charge. Such molecules generate a
new form of effective multi-fermion effective interaction similar
to the orignal NJL model. Brown et al.\cite{bglr2003} (denoted as
BGLR below) obtained the interaction induced by the instanton
molecules above $T_c$ by continuing the Nambu-Jona-Lasinio
description upwards from below $T_c$.

Our present discussion
of mesonic bound states  should not be confused with
quasi-hadronic states  found in early lattice
calculations\cite{detar85} for quarks and antiquarks propagating
in the space-like direction. Their spectrum, known as ``screening
masses'' is generated mostly by ``dynamical confinement" of the
spatial Wilson loop which is a nonperturbative phenomenon seen via
the lattice calculations. Similar effects will be given here by
the instanton molecule interaction.


\section{Binding of the $\bar q q$ states}
\label{binding}
\subsection{The Coulomb interaction and the relativistic effects}
\label{sec_coul_rel}
At $T>T_c$ the charge is screened rather than confined \cite{Shu_JETP},
and so the potential has a general Debye form
\be V= {\alpha_s(r,T) \over r} exp\left(-{r\over R_D(T)}\right)\ee
(Note  that we use a (somewhat nonstandard)
 definition in which $\alpha_s$ absorbs the
  4/3 color factor.)
The general tenet of QCD tells us that the strength of the color
Coulomb should run. We know that perturbatively it should run as
\be
\alpha_s\sim\frac{1}{\log (Q/\Lambda_{\rm QCD})}
 \ee
with $\Lambda_{\rm QCD}\sim 0.25$ GeV. The issue is what
happens when the coupling is no longer small. In vacuum
we know that the electric field is ultimately confined to
a string, producing a linear potential.

In the so-called ``plasma phase" this does not happen, and SZ
assumed that the charge runs to larger values, which may explain
the weak binding at rather high T we discussed in the
introduction.
Lattice results produce potentials which, when fitted in the form
$V(r)=-A \exp(-mr)+B$ with constant $A,B$ indeed indicate that
$A(T)$ grows above $T_c$ by a large factor, before starting to
decrease logariphmically at high $T$. The maximal value of the
average coupling $max(A)\approx 1/2$. This is the value which will
keep charmonium bound, as found by Asakawa and Hatsuda, up to $1.6
T_c$\cite{hatsuda2003}.

Running of the coupling is
not very important for this work in which
 we are mostly interested in deeply bound states related with
 short enough distances.
Therefore we will simply keep it as a non-running constant, selecting some
appropriate average value.

It is well known in the point charge Coulomb problem (QED) that
when $Z\alpha$ is increased and the total energy reaches zero there is a
singularity,
preventing
solutions  for larger $Z\alpha$.
 In the problem of the ``sparking of the vacuum" in
relativistic heavy ion collisions, the
 solution of the problem
was found by approximating the nuclei by
 a uniformly charged sphere; for a review of the history
see Rafelski et al.~\cite{rafelski78}. As a result of such
 regularization,
the bound electron level continues  past zero to $-m$,
at which point $e^+ e^-$ production becomes possible
around the critical value of $Z_{cr}=169$.
 In short, the problem of the
point Coulomb charge could be taken care of by choosing a
distributed electric field which began from zero at the origin.

In QCD the charge at the origin is switched off
by asymptotic freedom, the coupling which runs to zero value at
the origin. A cloud of virtual fields making the charge
is thus ``empty inside''.
 We will  model a resulting potential
 for the color Coulomb interaction by simply
 setting the electric field equal to zero at $r=0$, letting it
decrease (increase in attraction) going outward.  We
can most simply do this by choosing a charge distribution which is
constant out to $R$, the radius of the meson. If the original
$2m_q$ mass were to be lowered to zero by the color Coulomb
interaction and instanton molecule interaction,
then the radius of the final molecule will be
\be
R\simeq\frac{\hbar}{2 m_q},
\ee
although the rms radius will be substantially greater
with the instanton molecule interactions playing the main role around $T_c$.
\be
V &=& 
      - \alpha_s\frac{1}{2R}\left(3-\frac{r^2}{R^2}\right),
      \;\;\;\;\; r<R \nonumber\\
   &=&  - \alpha_s\frac{1}{r}, \;\;\;\;\; r>R.\label{9}
\label{potential} \ee
This $V$ has the correct general
characteristics. As noted above, the electric field $\vec E$ must
be zero at $r=0$. It is also easy to see that $V$ must drop off as
$r^2/R^2$ as the two spheres corresponding to the quark and
antiquark wave functions are pulled apart. Precisely where the
potential begins the $1/r$ behavior may well depend upon
polarization effects of the charge, the $+$ and $-$ charges
attracting each other, but it will be somewhere between $R$ and
$2R$, since the undisturbed wave functions of quark and antiquark
cease to overlap here.

The $q\bar q$ system is similar to positronium in the equality of
masses of the two constituents. Since the main term value
is $m\alpha^2/4$, the $4$, rather than 2 in hydrogen, coming from the
reduced mass, one might think that the Coulomb, velocity-velocity
and other interactions would have to be attractive and 8 times
greater than this term value in order to bring the $2 m_q$ in
thermal masses to zero. However, this does not take into account
the increase in reduced mass with $\alpha$. Breit and Brown
\cite{Breit48} found an $\alpha^2/4$ increase in the reduced mass
with $\alpha$, or $25\%$ for $\alpha=1$, to that order. It should
be noted that in the Hund and Pilkuhn \cite{pilkuhn00}
prescription the reduced mass becomes $\mu=m_q^2/E$, which
increases as $E$ drops.

We first proceed to solve the Coulomb problem, noting that this gives us
the solution to compare with the quenched lattice gauge simulations,
which do not include quark loops.

Having laid out our procedure, we shall proceed with
approximations. First of all, we ignore spin effects in getting a
Klein-Gordon equation. The chirally restored one-body equation
which has now left-right mixing is
given by
 \be (p_0 +
\vec\alpha\cdot\vec p)\psi =0.\label{chidirac}
 \ee
Expressing $\psi$ in two-component wave functions $\Phi$ and
$\Psi$, one has \be
p_0\Phi &=& -\vec\sigma\cdot\vec p\Psi \nonumber\\
p_0\Psi &=& -\vec\sigma\cdot\vec p\Phi, \label{eq9} \ee giving the
chirally restored wave function on $\Psi$ \be \left(p_0
-\vec\sigma\cdot\vec p \frac{1}{p_0} \vec\sigma\cdot \vec p
\right)\Psi =0. \label{eq11}
 \ee
Here
 \be p_0=E_V=E+\alpha_s/r.
 \ee
Neglecting spin effects, $\vec\sigma\cdot\vec p$ commutes with
$p_0$, giving the Klein-Gordon equation $p_0^2 - \vec{p}^2=0$. We
now introduce the effective (thermal) mass, so that the equations
for quark and hole can be solved simultaneously following
\cite{pilkuhn00};
 \be
\left[ (\epsilon -V(r))^2-\mu^2 -\hat p^2 \right] \psi (r) =0
\label{klein} \ee where $\hat p$ is momentum operator, and the
reduced energy and mass are $ \epsilon = (E^2-m_1^2-m_2^2)/2E, \mu
= m_1 m_2 /E $ with $m_1=m_2=m_q$.

 Furthermore from eq.(\ref{eq9}),
 \be
\la \vec{\alpha}\ra=(\Psi^\dagger, \vec{\sigma}\Phi) +
(\Phi^\dagger,
\vec{\sigma}\Psi)=\frac{\vec{p}}{p_0}-\frac{i}{p_0}\la
[\vec{\sigma}\times\vec{p}]\ra.
 \ee
If $\vec{\sigma}$ is parallel to $\vec{p}$, as in states of good
helicity, the second term does
not contribute. From the chirally restored Dirac equation
(\ref{chidirac}), ignoring spin effects such as the spin-orbit
interaction which is zero in S-states we are considering, we find
$p_0^2=\vec{p}^2$.

Brown \cite{brown52} showed that in a stationary state the EM
interaction Hamiltonian between fermions is
 \be
H_{\rm int} =\frac{e^2}{r}\left(
1-\vec\alpha_1\cdot\vec\alpha_2\right),\label{hint}
 \ee
where the $\vec\alpha_{1,2}$ are the velocities. Applying
(\ref{hint}) to the chirally restored domain of QCD, we expect
 \be
H_{\rm int} &=& \frac{2\alpha_s}{r} \;\;\;\;\;\;\;\;{\rm for}\;\;
                            \vec\alpha_1\cdot\vec\alpha_2=-1\nonumber\\
            &=& 0  \;\;\;\;\;\;\;\;{\rm for}\;\;
  \vec\alpha_1\cdot\vec\alpha_2=+1
\label{eq14}
 \ee

\subsection{The spin-spin interaction}
\label{sec_spinspin}
  The nonrelativistic form of the spin-spin interaction,
in the delta-function form, may give an impression that it is
maximal at the smallest distances. However this is not true, as
becomes clear if the relativistic motion is included in full, and
in fact at $r\rightarrow 0$ it is suppressed. At large r,
when particle motion is slow, it is of course again suppressed,
thus contributing mostly at some intermediate distances.

This fact is clear already from the derivation of  the $1s$-state
hyperfine splitting Fermi-Breit due to hyperfine interaction in
hydrogen from 1930  \cite{fermi30} given by
 \be \delta H=\frac 23
(\vec\sigma\cdot\vec\mu) \int d^3r\frac{\psi^\dagger\psi}{r^2}
\frac{d}{dr}\frac{e}{E+e^2/r+m}. \label{eq2.2}
 \ee
Note the complete denominator, which non-relativistically is just
substituted by m alone, but in fact contains the potential and is
singular at $r\rightarrow 0$. The derivative of the $e^2/r$ in the
denominator insured that the electric field was zero at $r=0$.
Here $\vec\sigma$ is the electron spin, $\vec\mu$ the proton
magnetic moment. In eq~(\ref{eq2.2}) the derivative can then be
turned around to act on $\psi^\dagger\psi$, and to order
$\alpha=1$ and with the $e^2/r$ neglected in the denominator, one
has
 \be \delta H\simeq
-\frac{8\pi}{3} (\vec\sigma\cdot\vec\mu)\frac{e}{2m} \psi^2(0),
\label{eq6}
 \ee
with $\psi$ taken to be the nonrelativistic $1s$ wave function to
lowest order in $\alpha$.

The hyperfine structure is obtained by letting the first $\vec p$
in eq.~(\ref{eq11}) act on the $p_0^{-1}$ and the second
$\vec p$ go $\vec p +\sqrt{\alpha_s}\;\vec A$ with
\be
\vec A=\frac{\vec \mu\times
\vec r}{r^3}
\ee
with $\vec \mu$ the magnetic moment of the
antiquark. One finds that the hyperfine structure is
\cite{fermi30} \be H_{\rm hfs} =\frac{1}{p_0^2} \sqrt{\alpha_s}\;
\vec\sigma \cdot [\vec E\times\vec A] \ee where $\vec E$ is color
electric field.
Thus, \be H_{\rm hfs}
=\frac{\sqrt{\alpha_s}\;|\vec E|}{p_0^2}
\left(\frac{\vec\sigma\cdot\vec\mu}{r^2}
-\frac{\vec\sigma\cdot\vec r\; \vec\mu\cdot\vec r}{r^4}\right)
=\frac 23 \frac{\sqrt{\alpha_s}\;|\vec E|}{p_0^2}
\frac{\vec\sigma\cdot\vec\mu}{r^2}.
\ee where $|\vec E|=2\alpha_s/r^2$.
As in the hydrogen atom,
the magnetic moments of quarks and antiquarks are
 \be
\mu_{q,\bar q} = \mp\frac{\sqrt{\alpha_s}}{p_0+m_{q,\bar q}}
 \ee
except that the Dirac mass $m_{q,\bar q}=0$ and $p_0$, in which
the potential is increased by a factor of 2 to take into account
the velocity-velocity interaction, is now \be p_0 =E+2
(\alpha_s/r) \ee for QCD so that in terms of the quark and
antiquark magnetic moment operators,
 \be
H_{\rm hfs}=-\frac 23 \frac{|\vec E|}{p_0 r^2}
(\vec\mu_q\cdot\vec\mu_{\bar q}).
 \ee
Of course, our $p_0$ for the chirally restored regime has
substantial $r$ dependence, whereas the $e/r$ in the hydrogen atom
is generally neglected, and $E+m$ is taken to be $2m$, so that
$\mu_e=-e/2m_e$. From Fig.~\ref{pot} it will be seen that (square
of) the wave function is large just where $\alpha_s/r$ is large.

For rough estimates we use averages. We see that, as in Table~\ref{tab1},
if $E$ is to be
brought down by $\sim 0.5 m_q$
for the $\sigma$ and $\pi$  by the
Coulomb interaction, then \be
2\langle \alpha_s/r\rangle \simeq  \frac 12  m_q\simeq \frac 14 p_0 \ee
so that with
$\alpha_s\sim 0.5 $, \be \langle r^{-1}\rangle \simeq \frac 12 m_q. \ee
We next see that this is consistent with the spin splitting forming a
fine structure of the two groups, the lower lying $\sigma$ and
$\pi$, and the slightly higher lying vectors and axial vectors.
Using our above estimates, we obtain
 \be \langle H_{\rm
hfs}\rangle &\simeq& \frac{1}{24}\;\frac{1}{16} \vec\sigma_q\cdot\vec\sigma_{\bar q}
m_q. \label{hhfs}
 \ee
so that for the $\sigma$ and $\pi$ where
$\vec\sigma_1\cdot\vec\sigma_2=-3$ we have \be \langle
H_{hfs}\rangle \sim -\frac{m_q}{128}, \ee the approximate equality
holding when $\alpha_s= 0.5$. Note that the hyperfine effect is
negligible for the $\alpha_s \sim 0.5$.
Although formally eq.~(\ref{hhfs}) looks like the hyperfine
structure in the chirally broken sector, it is really completely
different in makeup.

In our expression for $\langle H_{hfs}\rangle$ we have the $r$
dependence as $(p_0 r)^{-4} r^{-1}$ and $p_0 r=4$, basically
because the Coulomb interaction lowers the $\pi$ and $\sigma$ only
$1/4$ of the way to zero mass. This explains most of the smallness
of the spin-dependent interaction.

A recently renewed discussion of spin-spin
and spin-orbit interactions in a
 relativistic bound states has been made by
Shuryak and Zahed \cite{SZ_spinorbit}, who derived their form for
both weak and  strong coupling limits. Curiously enough, the
spin-spin term changes sign between these two limits: perhaps this
is another reason why at intermediate coupling considered in this
work it happens to be so small.

\subsection{The resulting $\bar q q$ binding}
\label{sec_qqbin}

We first construct the bound states for $T\gsim T_c$,
at temperature close enough to $T_c$ so that we can take the
running coupling constants at $T=T_c+\epsilon$. The fact that we
are above $T_c$ is important, because the $\Lambda_{\rm\chi SB}
\sim 4\pi f_\pi\sim 1$ GeV which characterizes the broken symmetry
state below $T_c$ no longer sets the scale. Until we discover the
relevant variables above $T_c$ we are unable to find the scale
that sets $\alpha_s=\frac 43 \frac{g^2}{\hbar c}$, the color
Coulomb coupling constant.

 Following SZ~\cite{shuryak2003}, we adopt
quark-antiquark bound states to be the relevant variables and
specifically, the instanton molecule gas~\cite{SSV_95} as a
convenient framework. In particular, Adami et al.\cite{adami1991},
Koch and Brown\cite{koch1993}, and BGLR~\cite{bglr2003} have shown
that $\gsim 50\%$ of the gluon condensate is not melted at
$T=T_c$. The assumption motivated by Ilgenfritz \& Shuryak~\cite{IS} is
then that the glue that is left rearranges itself into gluon
molecules around $T=T_c$, i.e., what BGLR call ``epoxy". We have
quantitatively determined couplings for the mesons in the
instanton molecule gas by extending the lower energy NJL in the
chiral symmetry breaking region up through $T_c$~\cite{bglr2003}.
We set these couplings in order to fit Miller's \cite{miller00}
lattice gauge results for the melting of the soft glue.

\begin{figure}[ht]
\centerline{\epsfig{file=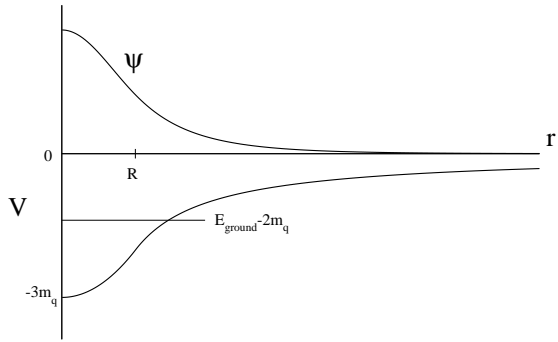,height=2.5in}}
\caption{The color Coulomb potential $V$ and
the corresponding wave function $\psi$ for
relativistic Klein-Gordon case. The interaction eq.~(\ref{potential})
with $R=\hbar/2 m_q$
was used.
The ground state energy with $\alpha_s=1$ corresponds to
$E_{\rm ground}=0.645\; m_q$. The minimum of the potential at the
origin is at $-3 m_q$ here.
}
\label{pot}
\end{figure}

In Fig.~\ref{pot} we show that if we choose $\alpha_s=1$
(effectively $\alpha_s=2$ by the doubling in Eq.~(\ref{eq14}))
as would be required to enter the strong coupling region
considered by Shuryak and Zahed\cite{shuryak2003b}
we bring the meson mass down by $-1.36\;
m_q$ from their unperturbed $2\; m_q$.
However, we switch to the region of $\alpha_s\sim 0.5$,
which is required by charmonium (intermediate coupling).
In Table~\ref{tab1} we summarize the Coulomb binding for a few choices
of $\alpha_s$.

 In the case of the instanton molecule interaction the
coupling constant $G=3.825$ GeV$^{-2}$ is dimensionful, so that
its contribution to the molecule energy scales as $G\; m_q^3$.
(Since we take $\alpha_s=0.5$ and will find that with inclusion
of the velocity-velocity interaction the effective $\alpha_s$ will
be 1, powers of $\alpha$ will not affect our answer. We will
use $m_q= 1$ GeV, essentially the lattice result for
$\frac 32 T_c$ and $3 T_c$\cite{petreczky02},
which works well in our schematic model.) Of course,
in QCD the Polyakov line goes to zero at $T_c$, indicating
an infinite quark mass below $T_c$; i.e., confinement.
Just at $T_c$ the logarithmically increasing confinement force
will not play much of a role because the dynamic confinement holds
the meson size to $\sim \hbar / m_q c$, or $\sim 0.2$ fm
with our assumption of $m_q =1$ GeV. (Later we shall see that
the rms radius is $\sim 0.3$ fm.) Since we normalize the instanton
molecule force, extrapolating it through $T_c$, and obtain the
color Coulomb force from charmonium, our $m_q$ is pretty well
determined. However, our $m_q=1$ GeV is for the unquenched system
and at a temperature where the instanton molecules play an important
role.

Given these caveats, we may still try to compare our Coulomb result
with the lowest peak of Asakawa et al.\cite{asakawa03} which is at
$\sim 2$ GeV for $T= 1.4 T_c \sim 0.38$ GeV and for Petreczky
at $\lsim 5 T\sim 2.030$ GeV for $T=1.5 T_c \sim 0.406$ GeV
where we used the Asakawa et al. $T_c$. We wish to note that:
{\it (i)} These temperatures are in the region of temperatures
estimated to be reached at RHIC, just following the color glass phase
(which is estimated to last $\sim 1/3$ fm/c). Indeed, Kolb et al.
begin hydrodynamics at $T=360$ MeV. {\it (ii)} These are in the region
of temperatures estimated by SZ\cite{shuryak2003} to be those for
which bound mesons form. We find these mesons to be basically
at zero binding, because the instanton molecule interactions although
important at $T=T_c$ (unquenched) because of the smallness ($\sim 1/3$ fm)
of the Coulomb $\bar q q$ states, will be unimportant at $T\sim 400$ MeV where
the molecules are much bigger. In the lattice calculations the scalar,
pseudoscalar, vector and axial-vector mesons come at the same energy.

Whereas there seems to be consistency between our estimates and
the giant resonances of both Asakawa et al.\cite{asakawa03}
and of Petreczky\cite{petreczky03}, we should note that with the
$m_q\sim 1.6$ GeV by Petreczky et al.\cite{petreczky02} the
mesons would still be bound by $\sim 1.2$ GeV at
$T=1.5 T_c$ (quenched). We do not think that the instanton molecules
should play an important role at such a high ($\sim 400$ MeV) temperature,
so this seems to be a discrepancy. Such a high binding would seem
to invalidate the SZ\cite{shuryak2003} need for the mesons to break
up around this temperature.
Earlier we have argued for a lower $m_q\sim 1$ GeV.

We are unable to extend our consideration to higher temperatures,
where the situation may move towards the perturbative one, but we
believe that lattice calculations do support our scenario that
the QGP contains large component of bound mesons  from
$T\sim 170$ MeV up to $T\sim 400$ MeV.

\begin{table}[ht]
\tbl{Binding energies from color Coulomb interaction
and the corresponding rms radii for various $\alpha_s$
(effectively, $2\alpha_s$ including velocity-velocity interaction).
4-point interactions are calculated using the parameters
obtained from color Coulomb interaction.
\label{tab1}
}
{\footnotesize
$
\begin{array}{cccc}
\hline
\phantom{xxx} \alpha_s \phantom{xxx} &  \Delta E_{\rm Coulomb}\; {\rm [GeV]} &
\phantom{xxx} \sqrt{\langle r^2\rangle} \; {\rm [fm]} \phantom{xxx} &
\Delta E_{\rm 4-point} {\rm [GeV]} \\
\hline
 0.50 &  - 0.483 &  0.360  &  - 0.994 \\
 0.55 &  - 0.595 &  0.313 &  - 1.385 \\
 0.60 &  -0.707 &   0.276 & - 1.834 \\
 1.00 & - 1.355 & 0.143 & - 7.574 \\
\hline
\end{array}
$
}
\end{table}

\begin{figure}[ht]
\centerline{\epsfig{file=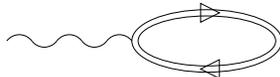,height=0.4in}}
\caption{Coulomb molecule. The wavy line on the left represents
the momentum transfer necessary to produce the molecule. The
double line denotes the Furry representation, i.e., Coulomb
eigenstate.} \label{fig3}
\end{figure}

For $\alpha_s=0.5$, which is the value required to bind charmonium
up through $T=1.6 T_c$, we find that the Coulomb interaction binds
the molecule by $\sim 0.5$ GeV, the instanton molecule interaction
by $\sim 1.5$ GeV. However, the finite size of the
$\psi^\dagger\psi$ of the instanton zero mode could cut the latter
down by an estimated $\sim 50\%$. As in the usual
NJL, there will be higher order bubbles, which couple the Coulomb
and instanton molecule effects.
We draw the Coulomb molecule in Fig.~\ref{fig3}, where the double
lines denote the Furry representation (Coulomb eigenfunction for
quark and antiquark in the molecule).

\begin{figure}[ht]
\centerline{\epsfig{file=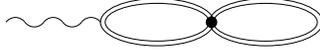,height=0.25in}}
\caption{The four-point instanton molecule interaction between Coulomb
eigenstates. The $(\bar\psi\psi)^2$ intersect at the thick point.
}
\label{fig4}
\end{figure}

\begin{figure}[ht]
\centerline{\epsfig{file=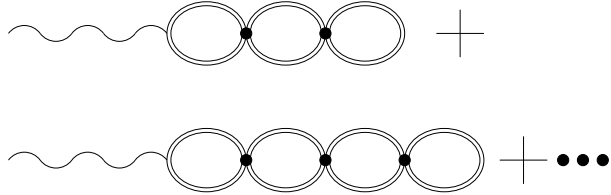,height=1.0in}}
\caption{Higher order effects of four-point interaction.}
\label{fig5}
\end{figure}

The four-point instanton molecule interaction is shown in Fig.~\ref{fig4}.
There will be higher-order effects as shown in Fig.~\ref{fig5}, of the
4-point interaction used in higher-order between Coulomb eigenstates
which always end in a 4-point interaction. The energy of the propagators
has been lowered from the 2 GeV of the two noninteracting quarks to
1.5 GeV by the Coulomb interaction. The series beginning with terms
in Figs.~\ref{fig3}$-$\ref{fig5} is
\be
\Delta E &=& -0.5 {\rm GeV} - 1 {\rm GeV}\; F
  -\frac{1 ({\rm GeV})^2 F^2}{1.5 {\rm GeV}}
  -\frac{1 ({\rm GeV})^3 F^3}{(1.5 {\rm GeV})^2} + \cdots
  \nonumber\\
&=& -0.5 {\rm GeV} -\frac{1 {\rm GeV} F}{1-\frac{1 {\rm GeV} F}{1.5 {\rm GeV}}}.
\label{eq43}
\ee
Now $\Delta E=-1.25$ GeV is accomplished for $F=0.5$.

Working in the Furry representation, we have a $-0.5$ GeV shift
already in the representation from the Coulomb wave functions.
This means that we must obtain $\Delta E=-1.5$ GeV to compensate
for the $2 m_q=2$ GeV, in order to bring the $\pi$ and $\sigma$
masses to zero. The four-point interaction is a constant, at
a given temperature, so this problem is just the extended
schematic model of nuclear vibrations (See Sec.~V of Brown
\cite{brown67}, where simple analytical solutions are given).

Our eq.~(\ref{eq43}) corresponds to the Tamm-Dancoff solution,
summing loops going only forward in time. If $\Delta E$ decreases
$-0.75$ GeV in this approximation, then when backward going graphs
are added \cite{BLRS}, $\Delta E$ will decrease by twice this
amount \cite{brown67}, or the $-1.5$ GeV necessary to bring the
$\pi$ and $\sigma$ energy to zero. Of course, forward and backward
going loops are summed in the Bethe-Salpeter equation to give the
NJL in the broken symmetry sector, but the actual summation is
more complicated there, because the intermediate state energies
are not degenerate. In the next section we shall show that the
backward going graphs appear in lattice gauge calculations.

In detail, with our estimated $F= (0.75)^2$ and the 4-point
energies from Table~\ref{tab1}, our $\pi$ and $\sigma$ excitations
without inclusion of backward going graphs are brought down $58\%$
of the way from $-0.5$ GeV to $-2$ GeV; i.e., slightly too far. We
have not made the adjustment down to $50\%$, because the
uncertainties in our estimate of $F$, etc., do not warrant greater
accuracy.


\section{Conclusions}

Shuryak and Zahed have discussed the formation of the mesonic
bound states at higher temperatures, well above $T_c$. They
pointed out that in the formation of the bound state, or any one
of the molecular excited states, the quark-quark scattering length
becomes infinite, similarly for the more strongly bound
gluon-gluon states. In this way the nearly instantaneous
equilibration found by RHIC can be explained. As we explained in
the last section, lattice calculations seem to support the
scenario of nearly bound scalar, pseudoscalar, vector and
axial-vector excitation at $\sim 2 T_c$ ($\sim 1.5$ times the
quenched $T_c$).

In this work we are able to construct a smooth transition from the chirally
broken to the chirally restored sector in terms of continuity in
the masses of the $\sigma$ and $\pi$ mesons, vanishing at
$T\rightarrow T_c$. In doing so we had to include relativistic
effects.
One of them -- the velocity-velocity term related to Amper law
for the interacting currents -- nearly doubles the effective coupling.
The spin-spin term happen to be very small. The crucial part
of strong binding in
our picture of $\bar q q$ mesons (or molecules) is
the quasi-local interaction due to instanton molecules
(the ``hard glue'').
We found that the tight binding of these mesons near $T_c$ enhences
the wave function at the origin, and gives us
additional understanding of the nonperturbative
hard glue (epoxy) which is preserved
at $T>T_c$.

Thus, we believe that the material formed in RHIC was at a
temperature where although the matter is formally in a quark-gluon
plasma phase, most of it is made of
chirally restored mesons.  Certainly this is not
the weakly coupled quark-gluon plasma expected at high $T$.

Finally, in this paper we have focused on quantum mechanical
binding effects in the vicinity of the critical temperature $T_c$
coming down from above.  Nice continuity in the spectra of the
light-quark hadrons -- e.g., the pions and the $\sigma$ -- across
the phase boundary should also hold for other excitations such as
the vector mesons $\rho,\omega, A_1$ which lie slightly above
$\pi$ and $\sigma$ because of quantum corrections. Since going
below $T_c$ from above involves a symmetry change from Wigner-Weyl
to Nambu-Goldstone, there is a phase transition and to address
this issue, it would be necessary to treat the four-fermi
interactions more carefully than in the pseudo-potential
approximation adopted here.  It seems plausible from the
renormalization group point of view~\cite{polchinski-shankar} that
the four-fermi interactions generated by the instanton molecules
-- attractive in all channels -- will not only trigger the quark
pairs to condense, thereby spontaneously breaking chiral symmetry
but also bring down the mass of the vector mesons, as the
temperature approaches $T_c$ from above. We will show in a future
publication~\cite{BLR04} how this phenomenon can take place in a
schematic model.

\section*{Acknowledgments}
GEB was partially supported by the
US Department of Energy under Grant No. DE-FG02-88ER40388.
CHL is supported by Korea Research Foundation
Grant (KRF-2002-070-C00027).



\begin{thebibliography}{99}

\bibitem{BLRS} G.E. Brown, C.-H. Lee, M. Rho, and E.V. Shuryak, hep-ph/0312175
{\it (denoted as BLRS)}.

\bibitem{BBP91} G.E. Brown, H.A. Bethe, and P.M. Pizzochero,
                Phys. Lett.  {\bf B263} (1991) 337.

\bibitem{BJBP93} G.E. Brown, A.D. Jackson, H.A. Bethe, and P.M. Pizzochero,
                Nucl. Phys. {\bf A560} (1993) 1035.

\bibitem{Hatsuda_sigma}T.~Hatsuda and T.~Kunihiro,
hep-ph/0010039; Phys. Lett. {\bf B145} (1984) 7;
Phys. Rev. Lett. 55 (1985) 158;
Prog. Theor. Phys. {\bf 74} (1985) 765.

\bibitem{SS_survive}
T.~Schafer and E.~V.~Shuryak,
Phys.\ Lett.\  {\bf B356} (1995) 147.

\bibitem{datta02} S. Datta, F. Karsch, P. Petreczky, and I. Wetzorke,
Nucl. Phys. Proc. Suppl. {\bf 119} (2003) 487.

\bibitem{hatsuda2003} M. Asakawa and T. Hatsuda,
``$J/\psi$ and $\eta_c$ in the deconfined plasma fom lattice QCD,"
hep-lat/0308034.

\bibitem{MS}
T.~Matsui and H.~Satz,
{Phys.\ Lett.\ } {\bf B178} (1986)  416.

\bibitem{hydro2001} D. Teaney, J. Lauret, and E. Shuryak,
                   Phys. Rev. Lett. {\bf 86} (2001) 4783,
see also nucl-th/0110037 (unpublished);
P.F. Kolb, P. Huovinen, U.W. Heinz, and H. Heiselberg,
                   Phys. Lett. {\bf B500} (2001) 232.

\bibitem{Teaney2003} D. Teaney, ``The effects of viscosity on spectra,
elliptic flow and HBT radii", nucl-th/0301099.

\bibitem{PSS}
G.~Policastro, D.~T.~Son and A.~O.~Starinets,
Phys.\ Rev.\ Lett.\  {\bf 87} (2001) 081601.

\bibitem{shuryak2003} E. Shuryak and I. Zahed, hep-ph/0307267.

\bibitem{Li6} K. M. O'Hara et al,
  Science {\bf 298} (2002) 2179;
T. Bourdel et al, Phys. Rev. Lett. {\bf 91} (2003) 020402.

\bibitem{petreczky02} P. Petreczky, F. Karsch, E. Laermann, S. Stickan,
I. Wetzorke, Nucl. Phys. Proc. Suppl. {\bf 106} (2002) 513;
hep-lat/0110111.

\bibitem{SZ_CFT}
E.~Shuryak and I.~Zahed,
hep-th/0308073, Phys. Rev. D, in press.


\bibitem{IS}
E.~M.~Ilgenfritz and E.~V.~Shuryak,
Phys.\ Lett.\ B {\bf 325} (1994) 263.

\bibitem{bglr2003} G.E. Brown, L. Grandchamp, C.-H. Lee, M. Rho,
hep-ph/0308147.

\bibitem{detar85} C. DeTar, Phys. Rev. {\bf D32} (1985) 276;
{\bf D37} (1987) 2378.


\bibitem{Shu_JETP} E V Shuryak, Zh.E.T.F {\bf 74} (1978) 408;
   Sov. Phys. JETP {\bf 47} (1978) 212

%
%
%
%
%
%
%
%
%

\bibitem{rafelski78} J. Rafelski, L.P. Fulcher, and A. Klein, Phys. Rept.
{\bf 38C} (1978) 227.

\bibitem{Breit48} G. Breit and G.E. Brown, Phys. Rev. {\bf 74} (1948) 1278.

\bibitem{pilkuhn00} V. Hund and H. Pilkuhn, J. Phys. {\bf B33} (2000) 1617.

\bibitem{brown52} G.E. Brown, Phil. Mag. {\bf 43} (1952) 467.

\bibitem{fermi30} E. Fermi, Z.f.Phys. {\bf 60} (1930) 320;
G. Breit \& W. Doermann, Phys. Rev. {\bf 36} (1930) 1732.

\bibitem{SZ_spinorbit} E.~V.~Shuryak and I.~Zahed, hep-th/0310031.

\bibitem{SSV_95}T.~Sch\"afer, E.~V.~Shuryak and J.~J.~M.~Verbaarschot,
Phys.\ Rev.\ D {\bf 51} (1995) 1267; Nucl. Phys., {\bf B412} (1994) 143.

\bibitem{adami1991} C. Adami, T. Hatsuda, and I. Zahed, Phys. Rev.
{\bf D43} (1991) 921.

\bibitem{koch1993} V. Koch and G.E. Brown, Nucl. Phys. {\bf A560} (1993) 345.

\bibitem{miller00} D.E. Miller, hep-ph/0008031.

\bibitem{shuryak2003b} E. Shuryak and I. Zahed, hep-th/0308073.

\bibitem{asakawa03} M. Asakawa, T. Hatsuda, and Y. Nakahara,
Nucl. Phys. {\bf A715} (2003) 863c.

\bibitem{petreczky03} P. Petreczky, hep-ph/0305189.

\bibitem{brown67} G.E. Brown, {\it "Unified Theory of Nucleon Models
and Forces"}, 1967, North Holland Pub. Co., Amsterdam.

%

\bibitem{polchinski-shankar}  R. Shankar, Rev.
Mod. Phys. {\bf 66} (1994) 129; J. Polchinski, {\it Recent
Directions in Particle Theory: From Superstrings and Black Holes
to the Standard Model}, edited by J. Harvey and J. Polchinski
(World Scientific, Singapore, 1994) p235-274.

\bibitem{BLR04} G.E. Brown, C.-H. Lee and M. Rho, ``Chemical
equilibration in relativistic heavy-ion collisions," to appear.


\end{thebibliography}
\end{document}